\documentclass[conference,10pt]{IEEEtran}
\IEEEoverridecommandlockouts
\usepackage{cite}
\usepackage{amsmath,amssymb,amsfonts}
\usepackage[algo2e,ruled,vlined]{algorithm2e}
\SetCommentSty{scriptsize}
\SetKwComment{tcc}{\{}{\}}
\SetKwIF{If}{ElseIf}{Else}{if}{}{else if}{else}{endif}
\SetKwFor{While}{while}{}{endw}
\SetKwFor{ForEach}{for each}{}{endfch}
\usepackage{graphicx}
\usepackage{textcomp}
\usepackage{xcolor}
\graphicspath{{figures/}}
\def\BibTeX{{\rm B\kern-.05em{\sc i\kern-.025em b}\kern-.08em
    T\kern-.1667em\lower.7ex\hbox{E}\kern-.125emX}}
    
\begin{document}

\title{Node-Aware Improvements to Allreduce\\
\thanks{
This research is part of the Blue Waters sustained-petascale computing project, which is supported by the National Science Foundation (awards OCI-0725070 and ACI-1238993) and the state of Illinois. Blue Waters is a joint effort of the University of Illinois at Urbana-Champaign and its National Center for Supercomputing Applications.  This material is based in part upon work supported by the Department of Energy, National Nuclear Security Administration, under Award Number DE-NA0002374. }
}

\author{
\IEEEauthorblockN{Amanda Bienz}
\IEEEauthorblockA{\textit{Department of Computer Science} \\
\textit{University of Illinois}\\
\textit{at Urbana-Champaign}\\
Urbana, Illinois \\
bienz2@illinois.edu}
\and
\IEEEauthorblockN{Luke N. Olson}
\IEEEauthorblockA{\textit{Department of Computer Science} \\
\textit{University of Illinois}\\
\textit{at Urbana-Champaign}\\
Urbana, Illinois \\
lukeo@illinois.edu}
\and
\IEEEauthorblockN{William D. Gropp}
\IEEEauthorblockA{\textit{Department of Computer Science} \\
\textit{University of Illinois}\\
\textit{at Urbana-Champaign}\\
Urbana, Illinois \\
wgropp@illinois.edu}
}

\maketitle

\begin{abstract}
    The \texttt{MPI\_Allreduce} collective operation is a core kernel of many parallel codebases, particularly for reductions over a single value per process.  The commonly used allreduce recursive-doubling algorithm obtains the lower bound message count, yielding optimality for small reduction sizes based on node-agnostic performance models.  However, this algorithm yields duplicate messages between sets of nodes.  Node-aware optimizations in MPICH remove duplicate messages through use of a single master process per node, yielding a large number of inactive processes at each inter-node step.  In this paper, we present an algorithm that uses the multiple processes available per node to reduce the maximum number of inter-node messages communicated by a single process, improving the performance of allreduce operations, particularly for small message sizes.
\end{abstract}

\begin{IEEEkeywords}
Parallel, Parallel algorithms, Interprocessor communications
\end{IEEEkeywords}

\section{Introduction}
The advance of parallel computers towards exascale motivates the need for increasingly scalable algorithms.  Emerging architectures provide increased process counts, yielding the potential to run increasingly large and complex applications, such as those relying on linear system solvers or neural networks.  As applications are scaled to a larger number of processes, MPI communication becomes a dominant factor of the overall cost. 

\begin{figure}[ht!]
    \centering
    \includegraphics[width=0.4\textwidth]{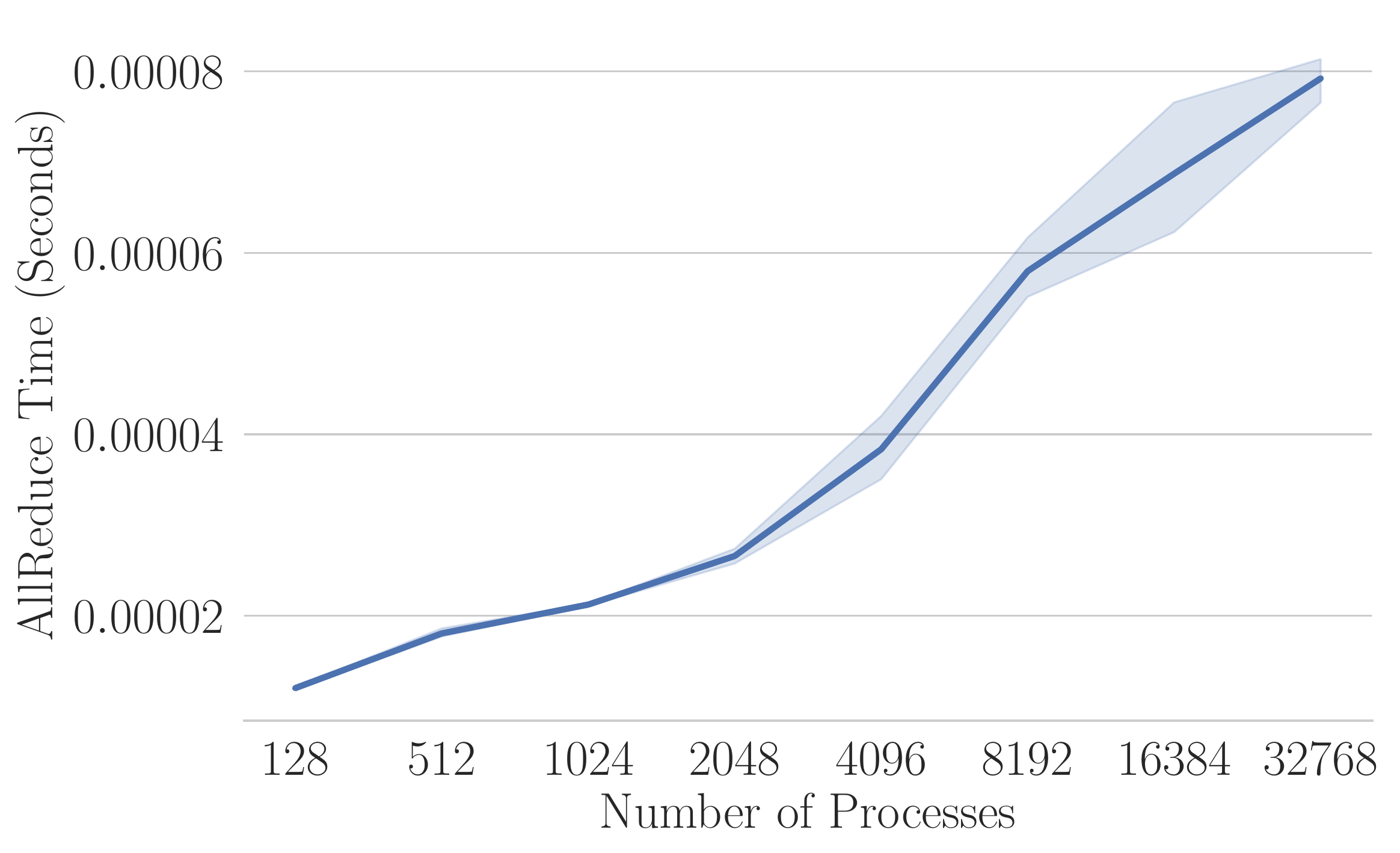}
    \caption{Time required for $\texttt{MPI\_Allreduce}$ to reduce a single double-precision floating-point value across a variety of process counts on Blue Waters~\cite{BlueWaters,bw-in-vetter13}.  The shaded area shows the variation between five separate runs.}\label{figure:scaling}
\end{figure}

The \texttt{MPI\_Allreduce}~\cite{MPI} is a fundamental component of a wide range of parallel applications, such as norm calculations in iterative methods, inner products in Krylov subspace methods, and gradient mean calculation in deep neural networks.  The allreduce operation consists of performing a reduction operation over values from all processes, such as a summing values or determining the maximum.  Therefore, the cost of the allreduce increases with process count, as displayed in Figure~\ref{figure:scaling}, motivating the need for improved performance and scalability on emerging architectures.  

In this paper, we present an allreduce algorithm based on node-awareness, which exchanges inter-node communication for less costly intra-node messages as well as increased computational requirements.  This algorithm reduces the number of inter-node messages from $\log_{2}(n)$ to $\log_{\texttt{ppn}}(n)$, where $n$ is the number of nodes involved and $\texttt{ppn}$ is the number of processes per node, yielding significant speedups over standard allreduce methods for small message sizes.

\begin{figure*}[t!]
    \centering
    \includegraphics[width=0.7\linewidth]{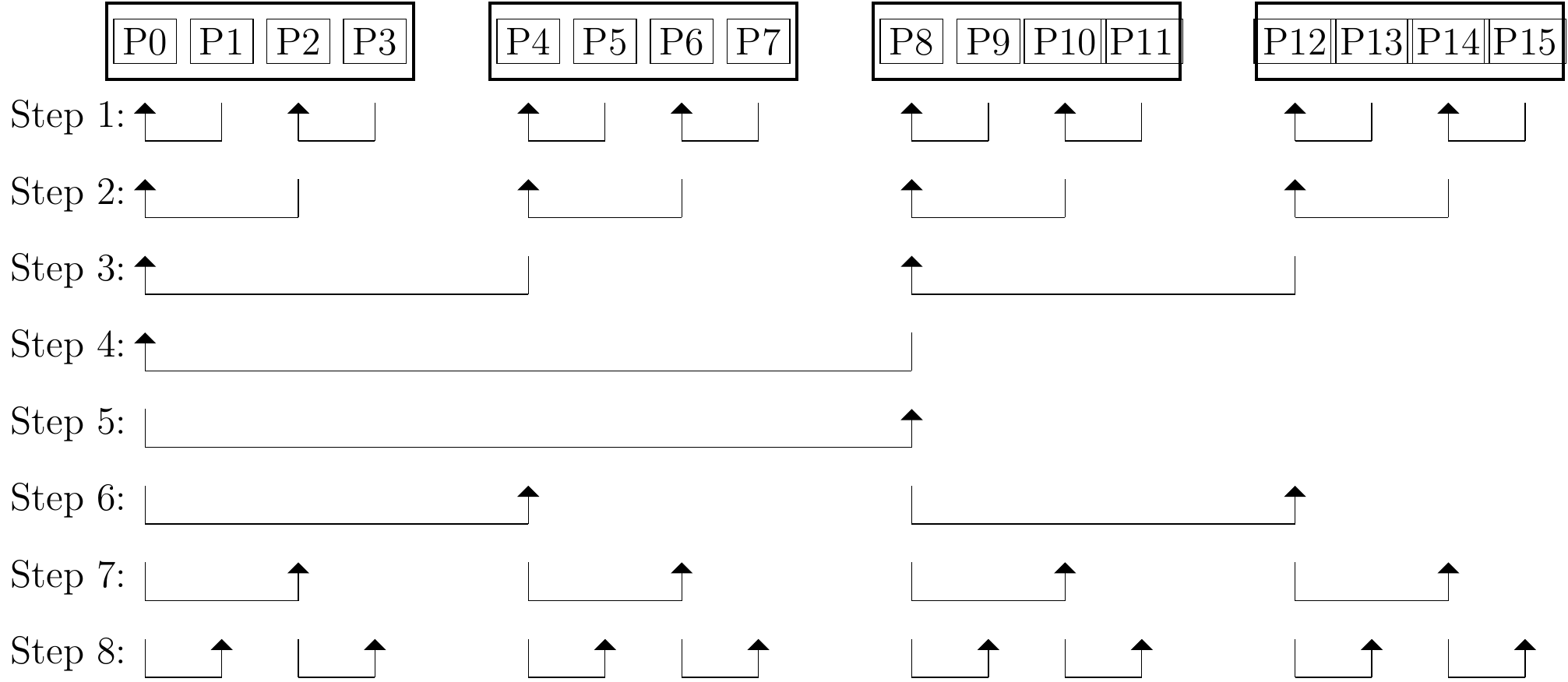}
    \caption{Data movement for a tree allreduce over $16$ processes, with data first reduced to a process $P0$ before being broadcast to other processes.}\label{figure:reduce_broadcast}
\end{figure*}
\begin{figure*}[t!]
    \centering
    \includegraphics[width=0.7\linewidth]{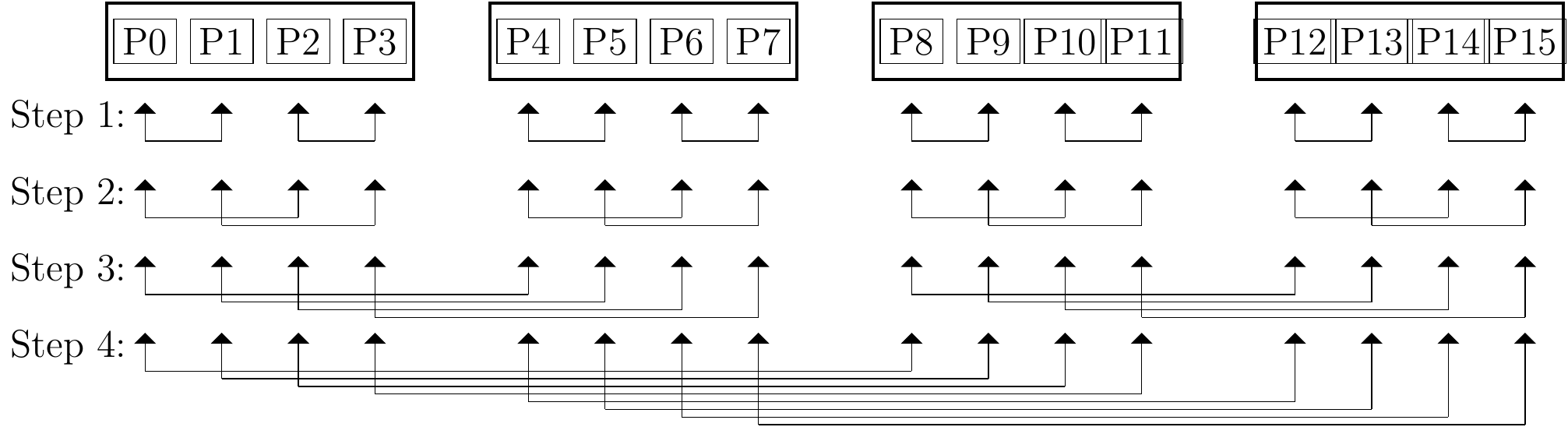}
    \caption{Communication pattern for a recursive-doubling allreduce with $4$ nodes, each containing $4$ processes.  Data is exchanged at each step and all processes are active in the reduction.}\label{figure:recursive_doubling}
\end{figure*}
\begin{figure*}[ht!]
    \centering
    \includegraphics[width=0.7\linewidth]{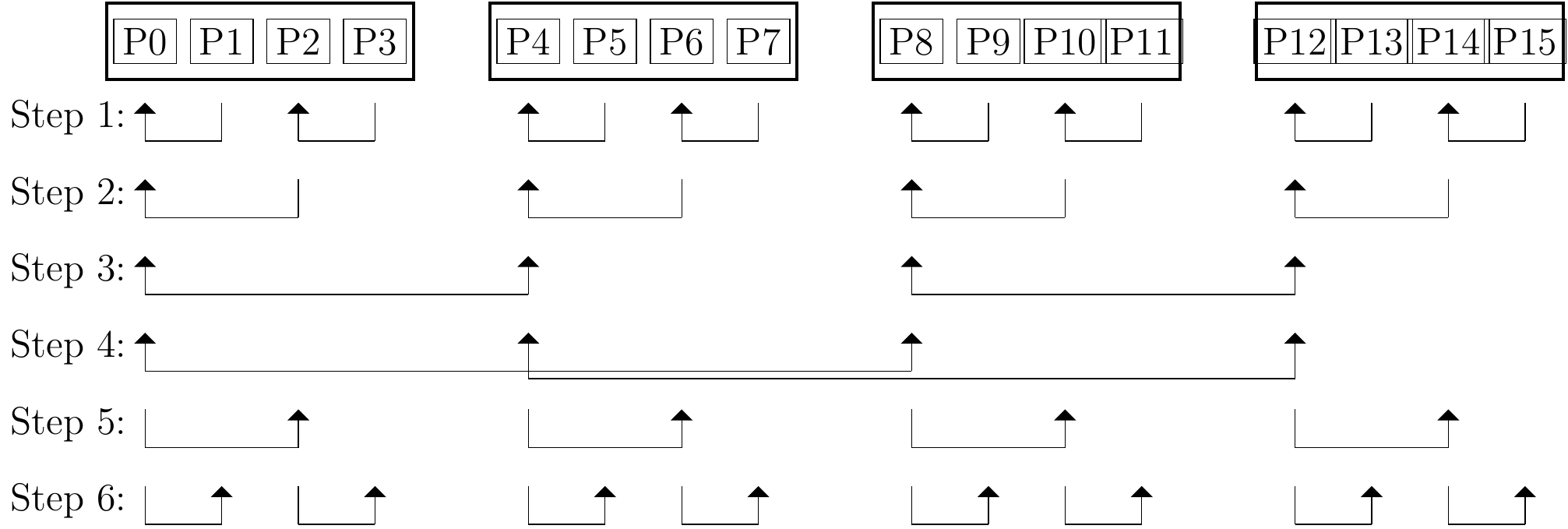}
    \caption{Data movement for the MPICH SMP allreduce algorithm over $16$ processes partitioned across $4$ nodes.  The data is reduced to a master process per node in steps 1 and 2, before being reduced among master processes through recursive doubling in steps 3 and 4.  Finally, the data is broadcast from the master process to all idle processes per node.}\label{figure:smp}
\end{figure*}
The remainder of this paper is organized as follows.  Section 2 describes common allreduce algorithms along with optimizations, including the node-aware allreduce algorithm that is implemented in MPICH~\cite{MPICH}.  In Section 3, we present a node-aware allreduce algorithm that reduces the number and size of inter-node messages.  Performance models for the various allreduce algorithms are analyzed in Section 4, and performance results are displayed in Section 5.  Finally, Section 6 contains concluding remarks.

\section{Background}\label{section:background}
The \texttt{MPI\_Allreduce} operates upon $s$ sets of $p$ values into $s$ resulting values through operations such as summations or calculating the maximum value.  These values are initially distributed evenly across $p$ processes and results are returned to all processes.  A reduction requires $(p-1)\cdot s$ floating-point operations if the full reduction is performed on a single process.  Therefore, splitting across $p$ processes yields a lower bound of $\frac{(p-1)\cdot s}{p}$ floating-point operations.  Furthermore, as data is distributed across all processes, a minimum of $\log_{2}(p)$ messages must be communicated.  Finally, the minimal data transfer size is $\frac{2\cdot (p-1) \cdot s}{p}$ as $\frac{(p-1) \cdot s}{p}$ values must be both sent and received~\cite{OptimizationMPICH,ChanTheory}.  

There are a large number of existing allreduce algorithms with various levels of optimality dependent on message size $s$ and process count $p$.  A straightforward algorithm, displayed in Figure~\ref{figure:reduce_broadcast}, first reduces the data onto a master process before broadcasting to all other processes.  

Assuming tree broadcasts and reductions, this algorithm requires $2 \cdot \log_{2}(p)$ messages and $2 \cdot \log_{2}(p) \cdot s$ values to be transported.  Furthermore, the tree algorithm requires $\log_{2}(p) \cdot s$ floating-point operations.  This algorithm is sub-optimal, with communication requirements significantly larger than ideal~\cite{Demystify}.  Furthermore, the tree algorithm yields large load imbalances with large numbers of inactive processes.

Recursive-doubling or the butterfly allreduce, exemplified in Figure~\ref{figure:recursive_doubling}, improves upon the tree algorithm by utilizing all processes, with sets of processes exchanging data at each step.  
This algorithm reduces the number and size of messages to $\log_{2}(p)$ and $\log_{2}(p) \cdot s$, respectively, while retaining computation requirements equivalent to the tree algorithm.  Recursive-doubling achieves the lower bound for message count, yielding a near-optimal algorithm, based on the postal model, for small messages and power of $2$ process counts~\cite{Demystify, Rabenseifner}.  This algorithm can be altered to work efficiently for non-power of two process counts~\cite{RecursiveMult, OptimizationMPICH}.

Alternative algorithms optimize bandwidth and local computation requirements for larger message sizes.  Assuming $p$ is relatively small, data can be split into $p$ portions and communicated to all other processes in a pipeline~\cite{PipelineOverlap}.  Portioning and pipelining the data achieves the lower bound cost associated with data transport.  However, each process sends $2 \cdot (p-1)$ messages, yielding reduced scalability for large process counts.  Rabenseifner's algorithm~\cite{Rabenseifner, OptimizationMPICH} improves upon pipelining by implementing a reduce-scatter, or a reduction with results scattered among the processes, followed by an allgather of these results.  While remaining optimal in data transport, this algorithm requires only $2\log_{2}(p)$ messages. 

\subsection{Node-Awareness}
Emerging architectures often consist of a large number of symmetric multiprocessing (SMP) nodes, each containing many processes.  Intra-node processes share memory, allowing for data to be quickly transported between processes on a node.  Inter-node data transport requires data to be split into packets, injected into the network, and transported across network links to the node of destination.  Therefore, inter-node communication is significantly more expensive than intra-node.  Node-agnostic performance models, such as the postal model, fail to accurately capture the costs associated with inter-node communication.  This model can be improved by splitting communication into intra- and inter-node as well as adding injection bandwidth limits~\cite{BienzEuroMPI, MaxRate}.

Standard allreduce methods, such as recursive-doubling and Rabenseifner's algorithm, reduce among processes in a node-agnostic fashion.  Therefore, multiple messages and duplicate data are often exchanged between a set of nodes.  This is exemplified in steps $3$ and $4$ of Figure~\ref{figure:recursive_doubling}, during which every process of one node is exchanging data with every process of another node, even though all processes per node hold identical values.

Duplicate inter-node communication can be removed through a node-aware SMP allreduce, displayed in Figure~\ref{figure:smp}, which reduces all intra-node data to a master process, performs a standard allreduce among master processes, and then broadcasts results locally~\cite{SMP}.  While the SMP approach requires the same number and size of inter-node messages as recursive-doubling, only a single process communicates from each node, eliminating injection bandwidth limits.  However, this approach yields a large number of inactive processes and load imbalance among processes on each node.

\subsection{Related Work}
Node-aware optimizations have been added to other collective algorithms~\cite{SMP ,HierAlltoAll}, with intra-node shared memory optimizations~\cite{HierSharedMem, HierSharedMemMultiCore}.  Similarly, node-awareness yields improvement to unstructured MPI communication, such as that which occurs during sparse matrix-vector multiplication~\cite{Bienz_napspmv}.  Furthermore, the order on which processes are mapped to each node can have a large effect on collective performance~\cite{DistanceAware, ProcessMapping}.  

Collective communication can be further improved through topology-awareness, reducing network contention and limiting message distance~\cite{TopoSubcommTaskMap,TopoNonMinimal,TopoMulticore, TopoInifiniband, TopoHierarchy}.  Collectives over large amounts of data can be further optimized for specific topologies~\cite{LargeCluster, LargeTreeTopo}.  Furthermore, as optimal algorithms depend on both message sizes as well as architectural topology, autotuners can determine the best algorithm for various scenarios~\cite{AutoTune, AutoTunedStarMPI}.  Finally, collective algorithms can be optimized for accelerated topologies, such as those containing Xeon Phi's~\cite{KNL} and GPU's\cite{GPUMVAPICH, GPUDirect, GPUEnergy, GPUIntraNode}.

\section{Node-Aware Parallel Allreduce (NAPAllreduce)}\label{section:napallreduce}
\begin{figure}[ht!]
    \centering
    \includegraphics[width=0.7\linewidth]{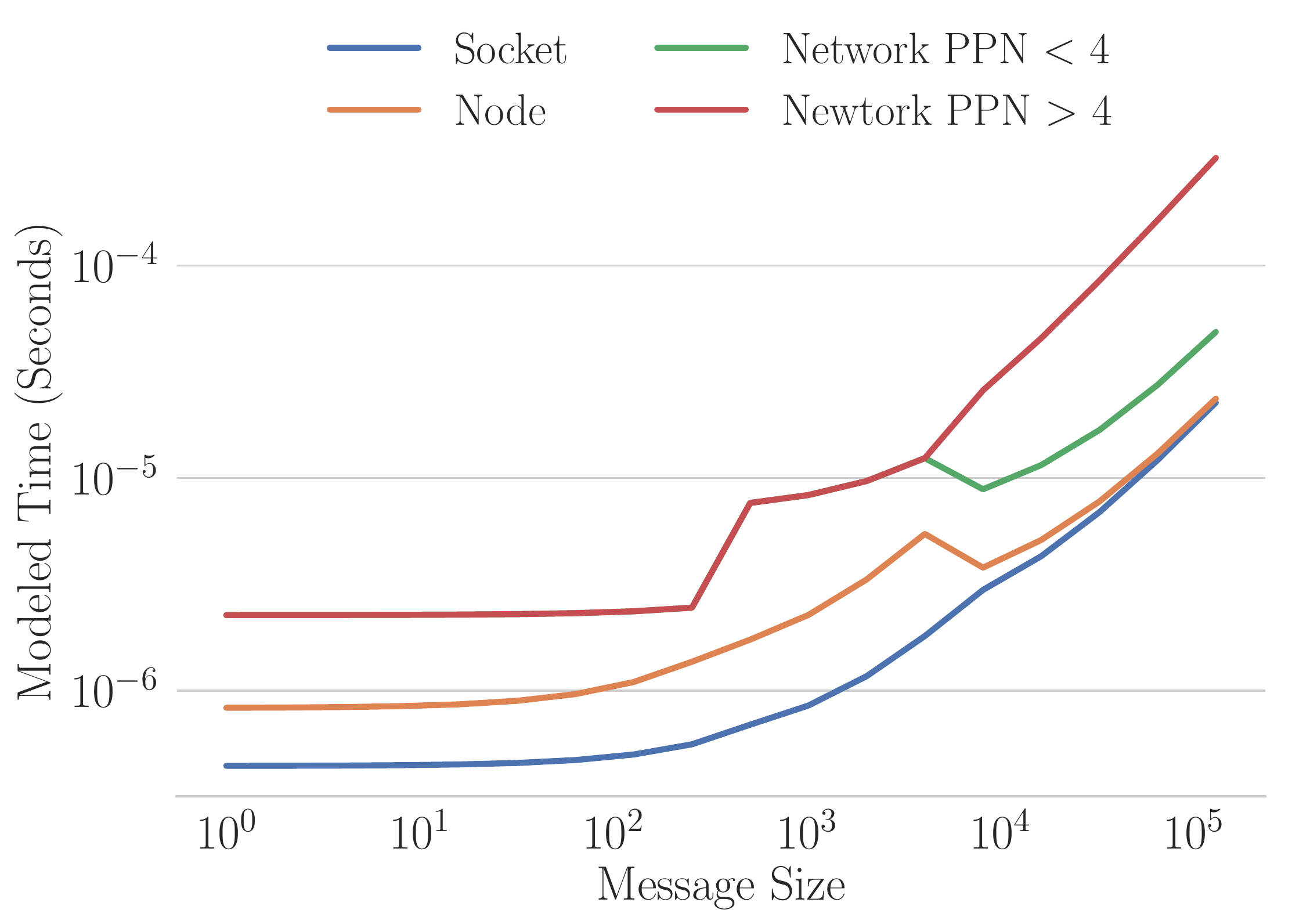}
    \caption{Modeled cost of communicating a single message between two processes on Blue Waters.  The costs are split into intra-socket ("Socket"), intra-node ("Node") and inter-node ("Network").  Inter-node communication costs are further split by \texttt{ppn} due to injection bandwidth limits.  These costs are calculated with the max-rate model.}\label{figure:max_rate}
\end{figure}

\begin{algorithm2e*}[ht!]
    \DontPrintSemicolon    \KwIn{data\tcc*{Data to be reduced}\\
          count\tcc*{Size of data}\\
          datatype\tcc*{MPI Datatype}\\
          MPI\_Op\tcc*{Reduction operation}
          comm, rank, num\_procs\tcc*{MPI Communicator for Allreduce}\\
          local\_comm, local\_rank, ppn\tcc*{Intra-node communicator}\\}
    \BlankLine    \KwOut{        reduced\_data\tcc*{Reduction of data over all processes in comm}
    }
    \BlankLine    
    \texttt{MPI\_Allreduce(data, reduced\_data, count, datatype, MPI\_Op, local\_comm)}\\
    \BlankLine    prev\_pos = 0\\
    subgroup\_size = ppn\\
    group\_size = subgroup\_size $\cdot$ ppn\\
    \For {i = 0 to $\log_{ppn}$($\frac{num\_procs}{ppn}$)}
    {
        group\_start = $\lfloor{\frac{rank}{group\_size}\rfloor}$ $\cdot$ group\_size\\
        subgroup = $\frac{rank \cdot group\_start}{subgroup\_size}$\\
        proc = group\_start $+$ prev\_pos $+$ local\_rank $\cdot$ subgroup\_size $+$ subgroup\\
        \If{$rank < proc$}
        {
            MPI\_Send(data, count, datatype, dest, tag, comm)\\
            MPI\_Recv(reduced\_data, count, datatype, dest, tag, comm, recv\_status)\\
        }
        \Else
        {
            MPI\_Recv(reduced\_data, count, datatype, dest, tag, comm, recv\_status)\\
            MPI\_Send(data, count, datatype, dest, tag, comm)\\
        }
        \texttt{MPI\_Op(reduced\_data)}\\
        prev\_pos = prev\_pos $+$ subgroup $\cdot$ subgroup\_size\\
        subgroup\_size = group\_size\\
        group\_size = group\_size $\cdot$ ppn\\
        \BlankLine        \texttt{MPI\_Allreduce(data, reduced\_data, count, datatype, MPI\_Op, local\_comm)}\\
    }
    
    \caption{NAP: \texttt{allreduce\_NAP}}\label{alg:allreduce_NAP}
\end{algorithm2e*}
The \texttt{MPI\_Allreduce} is commonly used for reductions on a small number of values per process, such as calculating an inner product of two vectors or a norm.  When reducing over a small set of values, the cost of the associated allreduce is dominated by the maximum number of messages communicated by any process.  Furthermore, the cost of each message is dependent on the relative locations of the sending and receiving processes.
Figure~\ref{figure:max_rate} displays the modeled cost of sending a single message containing of various sizes on Blue Waters, a Cray supercomputer at the National Center for Supercomputing Applications~\cite{BlueWaters, bw-in-vetter13}.  The costs were calculated with the max-rate model using parameters measured through ping-pong tests~\cite{MaxRate, BienzEuroMPI}.
Intra-socket messages, transported through cache, are significantly cheaper than inter-socket messages, which are transferred through shared memory.  Furthermore, inter-node communication is notably more expensive than intra-node.

Recursive-doubling requires each process on a node to communicate duplicate data at every inter-node step, yielding $\log_{2}(n)$ inter-node messages per process, where $n$ is the number of nodes.  The existing node-aware SMP algorithm improves the cost of relatively large reductions by removing duplicate messages between nodes, improving bandwidth costs.  However, the majority of processes remain idle as a single process per node performs all inter-node allreduce operations, requiring each master process to communication $\log_{2}(n)$ inter-node messages.  As a result, the maximum number of inter-node messages sent by a single process remains equivalent to recursive-doubling.  The remainder of this section introduces a node-aware allreduce algorithm, optimized for small reduction sizes, minimizing the maximum number of inter-node messages communicated by any process.

\begin{figure*}[ht!]
    \centering
    \includegraphics[width=0.7\linewidth]{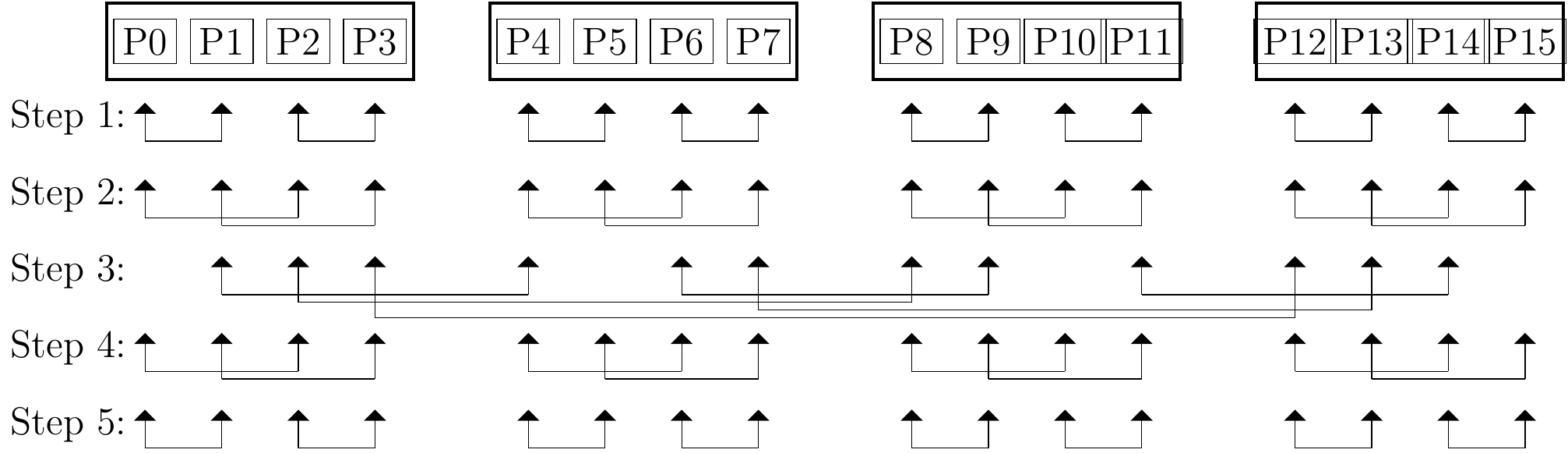}
    \caption{Communication pattern for the NAP allreduce method.  An intra-node allreduce is displayed in steps 1 and 2, while the single inter-node step is displayed in step 3.  Steps 4 and 5 consist of the final intra-node allreduce.  Note, one process per node sits idle during inter-node communication.}\label{figure:nap}
\end{figure*}

The SMP algorithm can be altered to use all \texttt{ppn} processes per node, splitting the required inter-node messages across all processes per node.  The node-aware parallel (NAP) method, exemplified in Figure~\ref{figure:nap}, consists of performing an intra-node allreduce so that all \texttt{ppn} processes hold a node's current reduction.  
Each process local to a node exchanges data with a specific node, before reducing results locally, as displayed in Step 3 of Figure~\ref{figure:nap}.  Therefore, the data from \texttt{ppn} nodes is reduced after a single step of inter-node communication, reducing the maximum number of inter-node messages to $\log_{\texttt{ppn}}(n)$.  For example, a reduction over $16$ nodes with $16$ processes per node requires only a single inter-node step.  Similarly, a NAP allreduce among $4096$ nodes, with $16$ processes each, requires only three inter-node steps.

\begin{figure*}[ht!]
    \centering
    \includegraphics[width=0.7\linewidth]{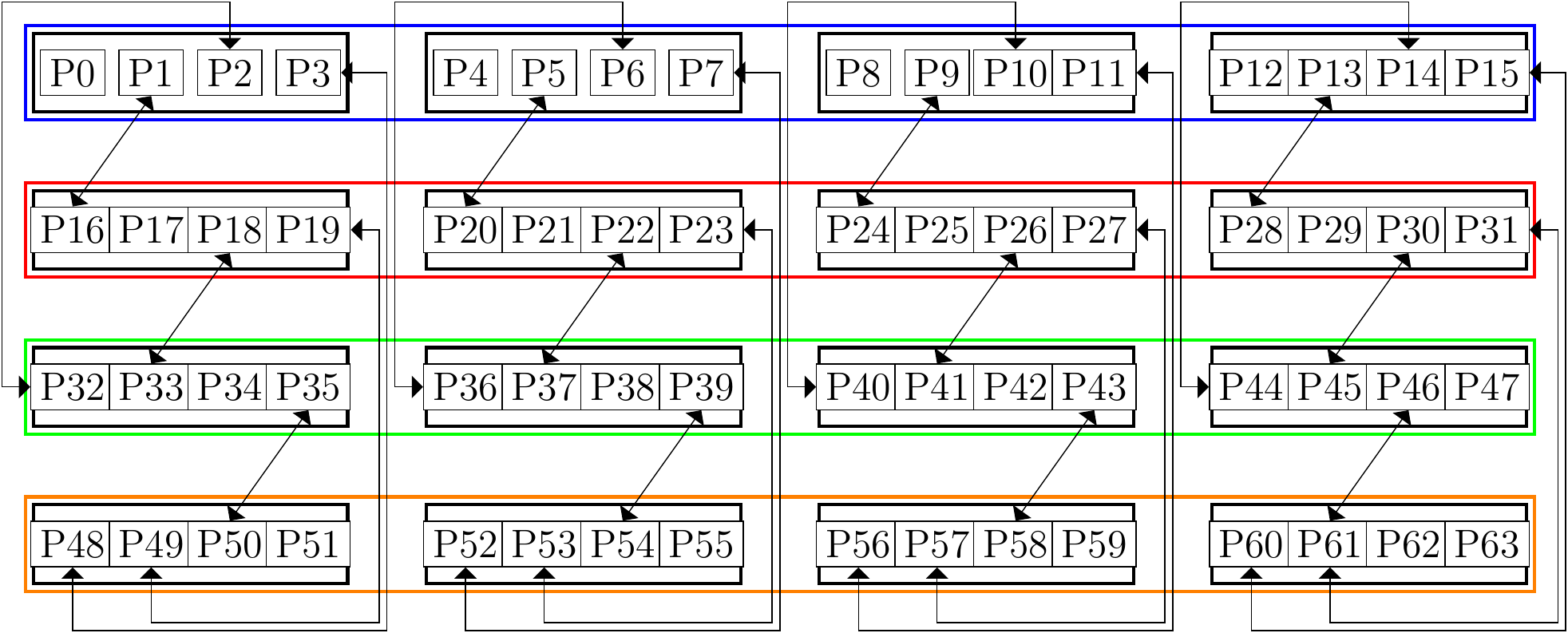}
    \caption{The second inter-node step of a NAP allreduce over $16$ nodes with $4$ processes per node.  Each row of nodes forms a subgroup, with all processes in a row containing equivalent data at the start of the step.}\label{figure:nap_step2}
\end{figure*}

The NAP allreduce algorithm is described in detail in Algorithm~\ref{alg:allreduce_NAP}.  At each inter-node step of this method, the number of nodes holding duplicate partial results increases by a power of \texttt{ppn}.  Initially, only processes on a single node hold equivalent reduction results.  However, at the beginning of the second inter-node step, all processes in a subgroup of \texttt{ppn} nodes hold equivalent data.  In general, at the start of the $i^{th}$ step, processes in each subgroup of $\texttt{ppn}^{i-1}$ nodes hold the same partial results.  Furthermore, a reduction is performed among groups of size $\texttt{ppn}^{i}$ at step $i$.  These groups and subgroups are exemplified in Figure~\ref{figure:nap_step2}, in which the second inter-node step of a NAP allreduce with $4$ processes per node is displayed.  
In the first step, each subgroup contains a single node and data is reduced over a row of nodes.  During the second step, each row of nodes forms a subgroup, with these subgroups outlined in color, and data is reduced among all $4$ subgroups.

Assuming SMP-style rank ordering, a process $q$ on node $m$ has local rank $r$, such that $q = n \cdot \texttt{ppn} + r$.  For example, process $P9$ in Figure~\ref{figure:nap} is located on node $2$ and has local rank $1$.  During each step of inter-node communication, process $q$ with local rank $r$ in subgroup $m$ communicates with process $u$ with local rank $m$ in subgroup $r$.  Therefore, process $P9$ from Figure~\ref{figure:nap} exchanges data with $P6$, which is located on node $1$ and has a local rank of $2$.  Note that any process with local rank equal to subgroup sits idly.

During later steps of communication, there are multiple processes with local rank $r$ in subgroup $m$.  Therefore, the node position, or the index of a rank's node within the subgroup, must remain constant.  In the second step of communication, displayed in Figure~\ref{figure:nap_step2}, process $P9$ with local rank $1$ in subgroup $0$ has node position $2$ as it lies on the third node in subgroup $0$. Therefore, $P9$ exchanges data with $P24$ as this process has local rank $0$ in subgroup $1$ and also has node position $2$.

\subsection {Non-Power of \texttt{ppn} Processes}
\begin{figure*}[ht!]
    \centering
    \includegraphics[width=0.7\linewidth]{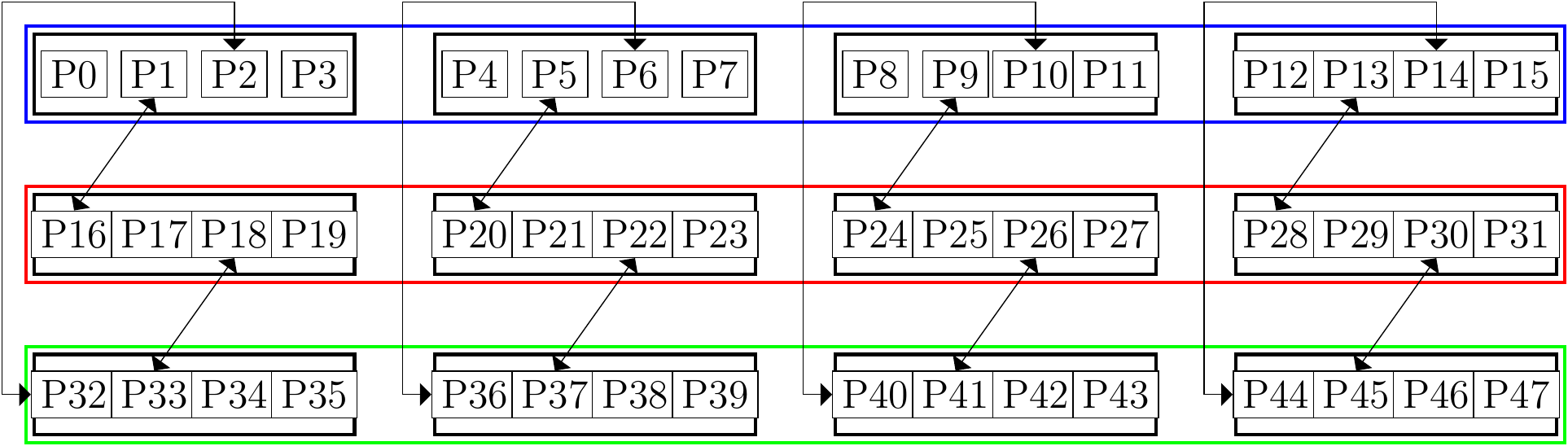}
    \caption{Communication pattern for a NAP allreduce with a non power of \texttt{ppn} process count.  As the number of nodes is divisible by \texttt{ppn}, the first step proceeds as normal, while the second step reduces over $3$ subgroups.  An extra process per node remains inactive for only the second step.}\label{figure:nap_power2}
\end{figure*}

The NAP allreduce algorithm reduces values among $p$ processes with only $\log_{\texttt{ppn}}(n)$ steps of inter-node communication.  However, this algorithm requires that the number of processes is a power of $\texttt{ppn}$, limiting process counts for which this algorithm is viable.  Assuming the number of nodes evenly divides \texttt{ppn}, the final step of inter-node communication can be reduced to involve only the necessary number of processes per node.  Figure~\ref{figure:nap_power2} displays the final step of a NAP allreduce with $12$ nodes and $4$ processes per node.  
All processes with local rank $3$ sit idly during the final step of inter-node communication as there are no available nodes with which to communicate.  However, the idle ranks recover the final result during the following intra-node allreduce.

\begin{figure*}[ht!]
    \centering
    \includegraphics[width=0.7\linewidth]{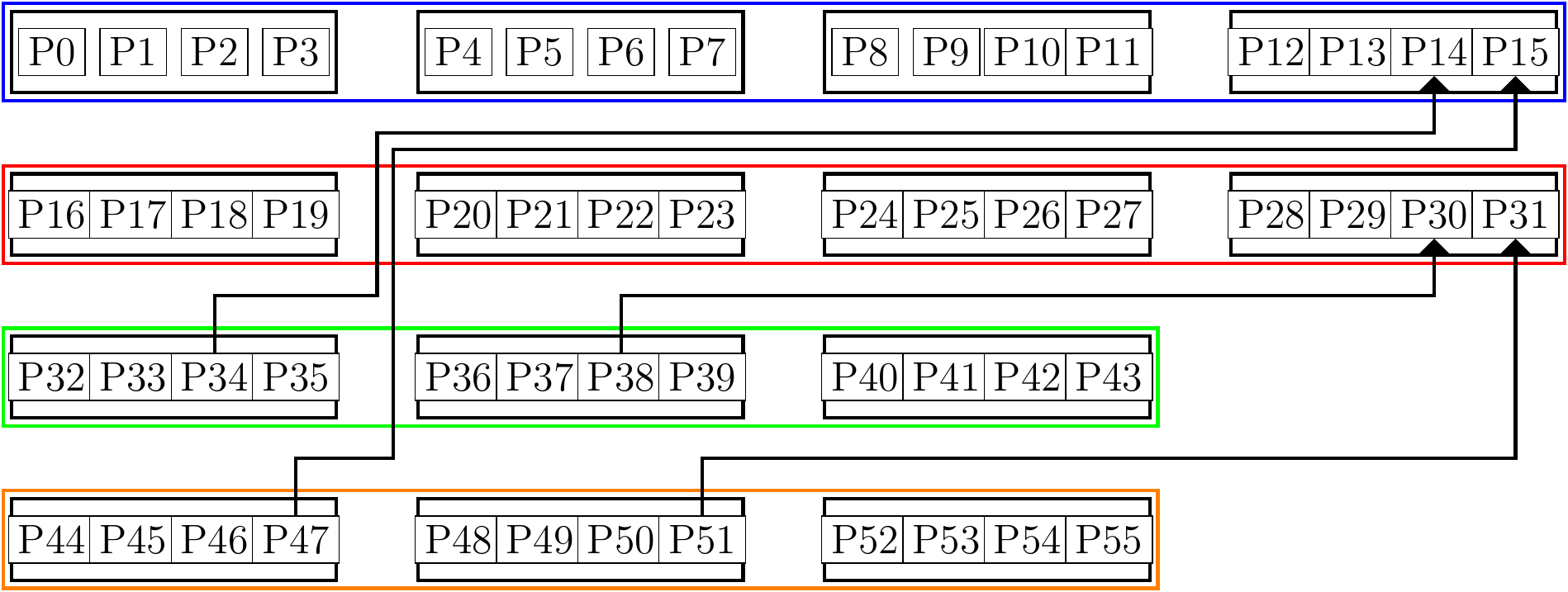}
    \caption{The communication pattern for a NAP allreduce with a number of nodes that does not divide \texttt{ppn}.  The initial step reduces over groups of nearly equal size.  The second inter-node step reduces as normal if the corresponding process exists, and otherwise receives data from the corresponding idle process.}\label{figure:nap_notdiv}
\end{figure*}

The NAP allreduce can also be extended to node counts that are not divisible by \texttt{ppn}.  In this case, subgroup sizes will not be equivalent during the final step of inter-node communication, as displayed in Figure~\ref{figure:nap_notdiv}.
Subgroups with extra nodes will have no corresponding process with which to reduce data, meaning some nodes will not achieve the full reduction.  However, as one process per node is idle during each step of inter-node communication, specifically the process with local rank equivalent to subgroup, each node has the potential to communicate with an extra node at each step.  Therefore, the processes on extra nodes that have no corresponding process with which to exchange will instead send data to the idle process.  Note this process does not need to receive data from the corresponding subgroup.  As an example, process $P14$ receives data from $P34$ during the final step of inter-node communication, as a corresponding node in subgroup $2$ does not exist.

\section{Node-Aware Performance Modeling}\label{section:perf_model}
Standard allreduce algorithms, such as recursive-doubling, minimize communication costs based on the standard postal model 
\begin{equation}
    T = \alpha t + \beta s + \gamma c,
\end{equation}
where $\alpha$ is the per-message start-up cost, $\beta$ is the per-byte transport cost, $\gamma$ is the flop rate, and $t$, $s$, and $c$ are the number of messages, bytes, and floating-point operations, respectively.  However, the cost of communication varies greatly with intra-node communication requiring significantly less cost than inter-node.  Therefore, the performance model can better capture cost by splitting up intra- and inter-node costs~\cite{BienzEuroMPI}, yielding
\begin{equation}
    T = \alpha_{\ell} t_{\ell} + \beta_{\ell} s_{\ell} + \alpha t + \beta s + \gamma c
    \label{eqn:split_model}
\end{equation}
where $\alpha_{\ell}$, $\beta_{\ell}$, $t_{\ell}$, and $s_{\ell}$ all represent intra-node communication while the remaining variables model inter-node communication and local computation.  Finally, inter-node bandwidth is greatly dependent on the number of processes communicating per node as injection bandwidth limits slow transport of large messages.  Therefore, this model is further improved by incorporating the max-rate model~\cite{MaxRate}
\begin{equation}
    T = \alpha_{\ell} t_{\ell} + \beta_{\ell} s_{\ell} + \alpha t + \frac{\texttt{ppn}\cdot s}{\min\left(R_{N}, \texttt{ppn}\cdot R_{b}\right)} + \gamma c
    \label{eqn:maxrate}
\end{equation}
where $R_{b}$ in inter-process bandwidth, or the inverse of $\beta$, and $R_{N}$ is injection bandwidth.  Note, this reduces to Equation~\ref{eqn:split_model} when inter-process bandwidth is achieved.

Performance costs of the various allreduce algorithms for small messages can be analyzed through the improved performance model in Equation~\ref{eqn:maxrate}.  The performance model cost of an allreduce of size $s$ over $p$ processes with recursive-doubling is displayed in Equation~\ref{eqn:recursive_double}.
\begin{multline}
    \left(\alpha_{\ell} + \beta_{\ell} s\right) \cdot \left(\log_{2}(\texttt{ppn})\right) \\
    + \left(\alpha + \frac{\texttt{ppn} \cdot s}{\min(R_{N}, \texttt{ppn} \cdot R_{b})}\right) \cdot \left(\log_{2}(n)\right) + \gamma s \cdot \left(\log_{2}(p)\right)\label{eqn:recursive_double}
\end{multline}
Recursive-doubling requires $\log_{2}(n)$ inter-node messages of size $s$, with injection bandwidth limiting performance for large values of $s$.  

The SMP allreduce improves upon this cost model, with the associated performance model cost of the SMP algorithm displayed in Equation~\ref{eqn:smp}.
\begin{multline}
    \left(\alpha_{\ell} + \beta_{\ell} s\right) \cdot \left(\log_{2}(\texttt{ppn})\right) \\
    + \left(\alpha + \frac{s}{R_{b}}\right) \cdot \left(\log_{2}(n)\right) + \gamma s \cdot \left(\log_{2}(p)\right)\label{eqn:smp}
\end{multline}
While the SMP method yields equivalent inter-node communication requirements to recursive-doubling, inter-node messages of all sizes achieve inter-process bandwidth as only a single process per node performs inter-node communication at any time.  The SMP approach does require slightly more intra-node communication than recursive-doubling due to the reduction and broadcast local to each node.

Finally, the NAP allreduce algorithm minimize inter-node communication requirements, exchanging inter-node messages for additional intra-node communication and local computation, as displayed in Equation~\ref{eqn:nap}.
\begin{multline}
    \left(\alpha_{\ell} + \beta_{\ell} s\right) \cdot \left(\log_{2}(p)\right) \\
    + \left(\alpha + \frac{\texttt{ppn} \cdot s}{\min(R_{N}, \texttt{ppn} \cdot R_{b})}\right) \cdot \left(\log_{\texttt{ppn}}(n)\right)\\
    + \gamma s \cdot \left(\log_{2}(p) + \log_{\texttt{ppn}}(n)\right)\label{eqn:nap}
\end{multline}
The number of inter-node communication steps is reduced from $\log_{2}(n)$ to $\log_{\texttt{ppn}}(n)$.  However, intra-node communication steps increase greatly from $\log_{2}(\texttt{ppn})$ to $\log_{2}(p)$ and and additional $\log_{\texttt{ppn}}(n)$ steps of local computation are required.  Furthermore, injection bandwidth will limit the rate at which bytes are transported for large messages as many processes per node are active in intra-node communication at each step.  Therefore, the NAP allreduce is ideal for small reduction sizes across a large number of processes, where extra computation and bandwidth injection limits are not a factor.

\begin{figure}[ht!]
    \centering
    \includegraphics[width=\linewidth]{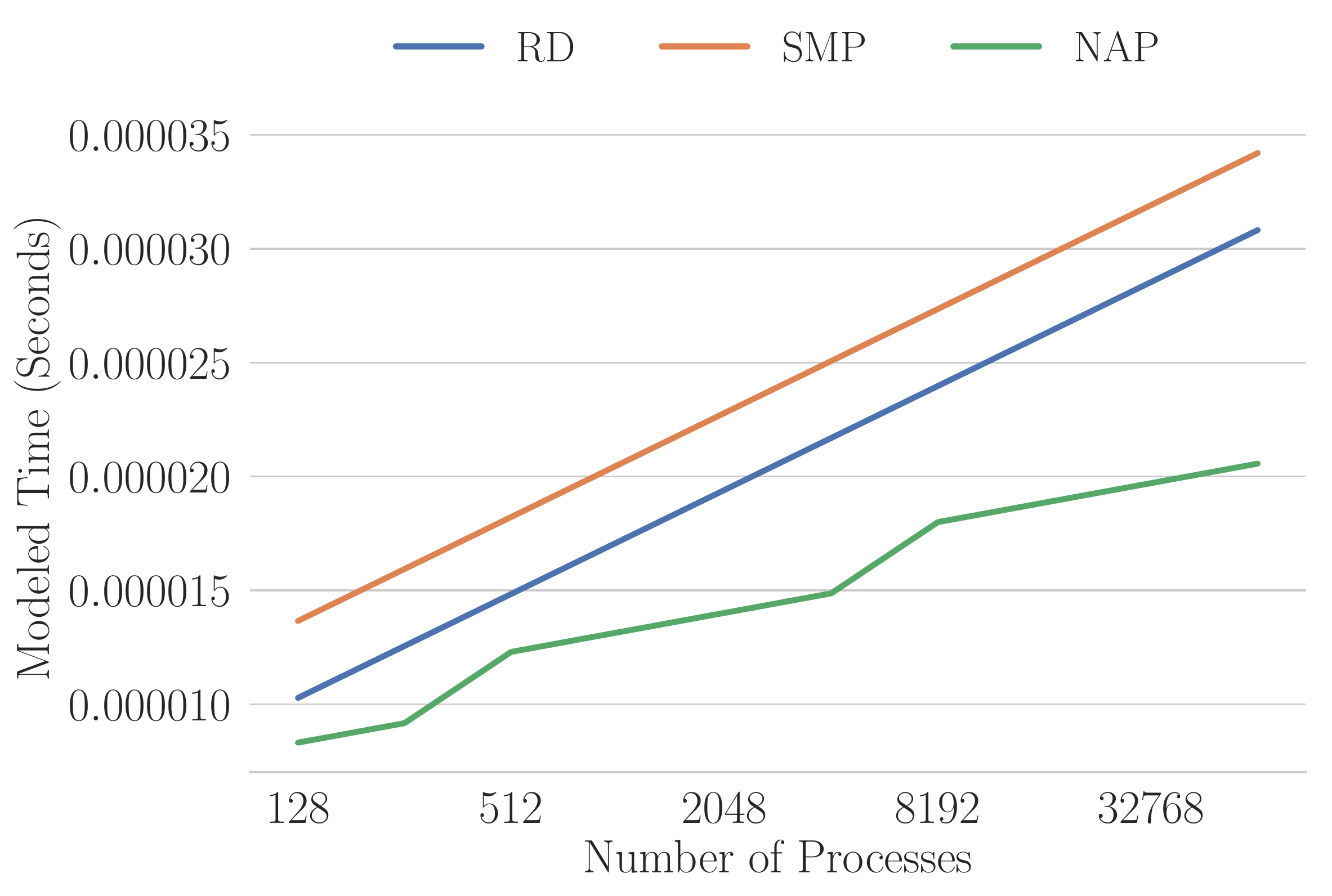}
    \caption{The modeled allreduce cost for reducing a single value across various process counts with the recursive-doubling (RD), SMP, and NAP methods.}\label{figure:perf_model_scale}
\end{figure}

\begin{figure}[ht!]
    \centering
    \includegraphics[width=\linewidth]{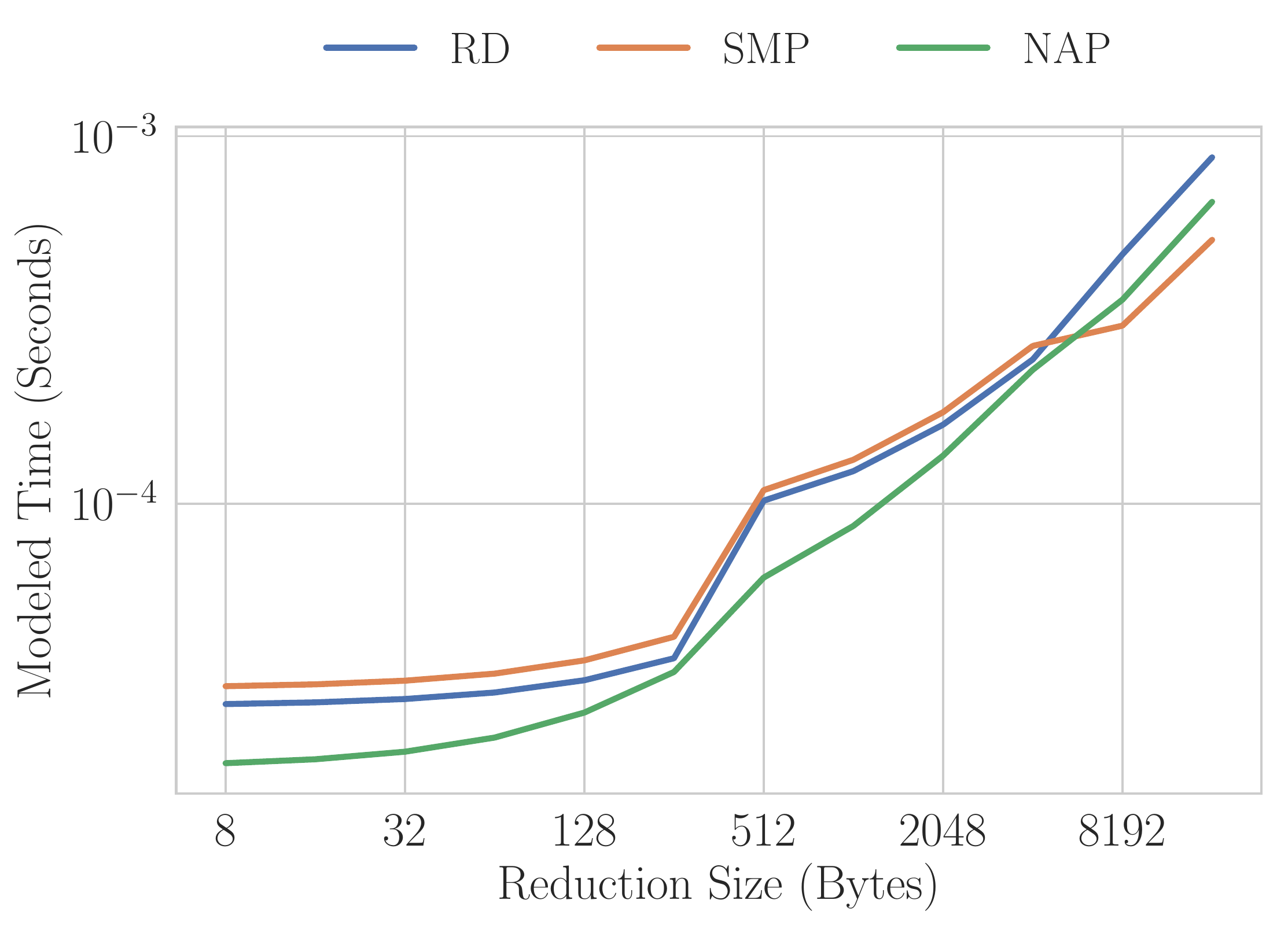}
    \caption{The modeled cost of performing an allreduce of various reduction sizes with each method on $32\,768$ processes.}\label{figure:perf_model_sizes}
\end{figure}

Figure~\ref{figure:perf_model_scale} shows the performance model costs for the recursive-doubling (RD), SMP, and NAP allreduce methods when reducing a single value across various process counts.  The model parameters were measured for Blue Waters with ping-pong tests and the STREAM benchmark~\cite{McCalpin1995, McCalpin2007}.
The performance models indicate that the NAP allreduce outperforms the other methods for small message sizes, particularly as process count increases.  Furthermore, Figure~\ref{figure:perf_model_sizes} displays the performance model costs for performing an allreduce with each method using $32,768$ processes, indicating the NAP allreduce outperforms recursive-doubling and SMP methods for small message sizes, while the SMP allreduce outperforms the recursive-doubling and NAP methods for large message sizes.

\section{Results}\label{section:results}
The recursive-doubling, SMP, and NAP allreduce algorithms were implemented on top of CrayMPI, utilizing the \texttt{MPI\_Send} and \texttt{MPI\_Recv} methods for each exchange of data.  Due to the associated overhead, results are presented for these implementations rather than comparing with recursive-doubling and SMP implementations that exist in MPICH.  All tests were performed on Blue Waters with $16$ processes per node.  Furthermore, each timing was calculated by performing thousands of allreduce operations to reduce error from timer precision, and each of these tests was performed $5$ times on different partitions of Blue Waters.  Each plot contains lines displaying the average results over the $5$ separate runs and outlines show the variation in timings over these $5$ tests.

\begin{figure}[ht!]
    \centering
    \includegraphics[width=\linewidth]{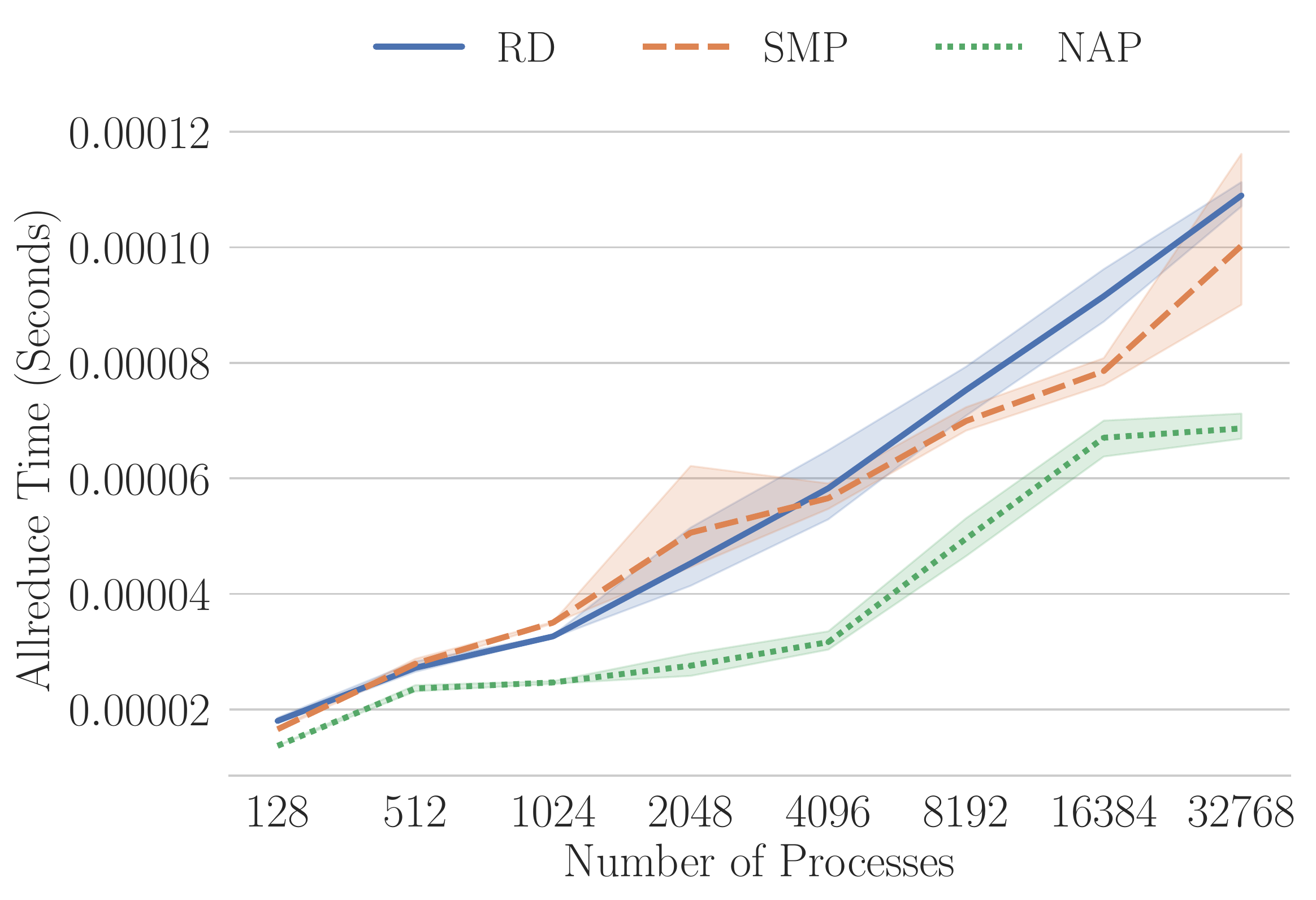}
    \caption{The measured cost of performing an allreduce of a single value with the recursive-doubling (RD), SMP, and NAP methods.}\label{figure:allreduce_scale_times}
\end{figure}

\begin{figure}[ht!]
    \includegraphics[width=\linewidth]{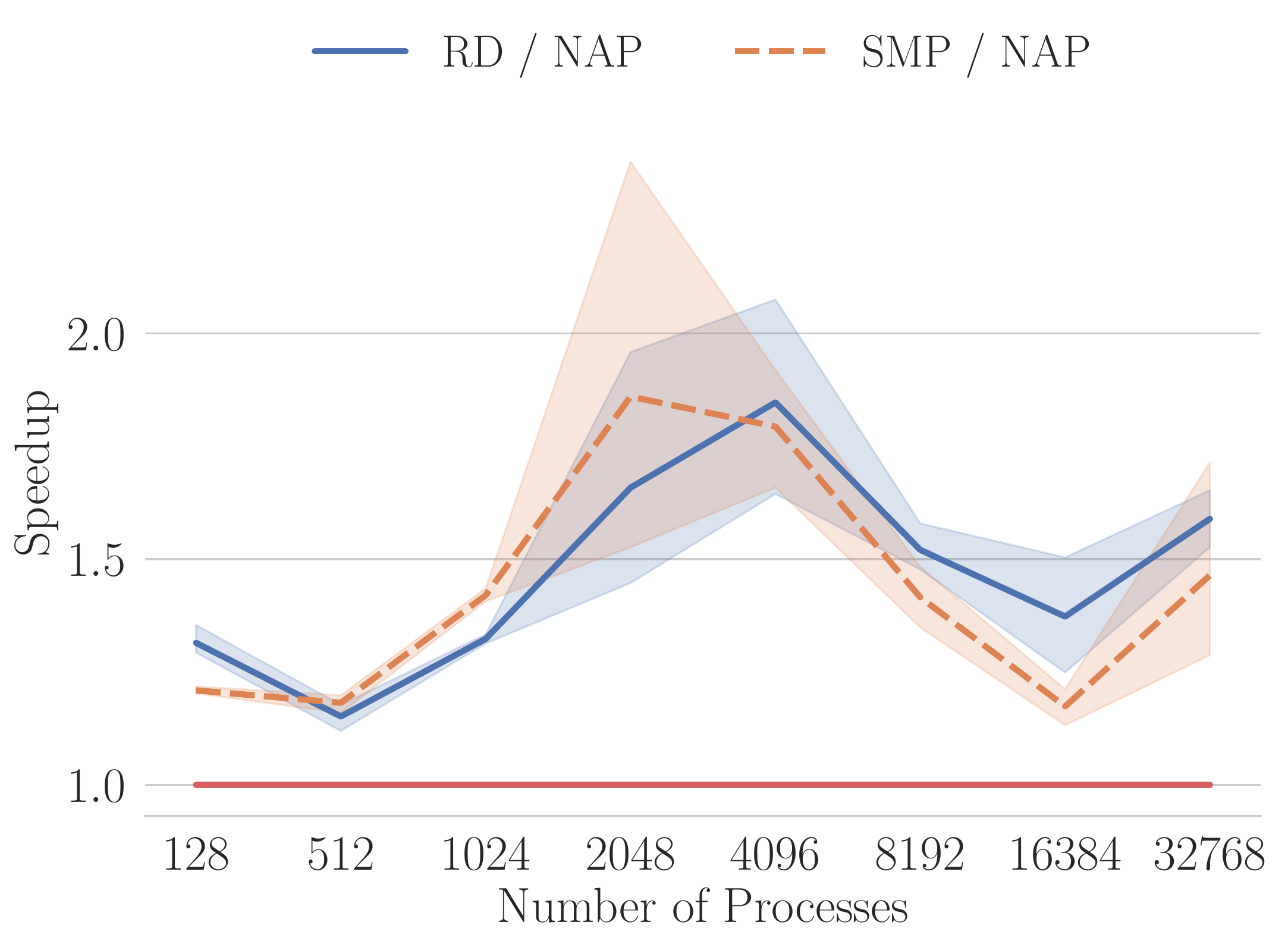}
    \caption{Speedup acquired from the NAP allreduce over the recursive-doubling and SMP methods when reducing a single value.}\label{figure:allreduce_scale_speedup}
\end{figure}

Figure~\ref{figure:allreduce_scale_times} displays the cost of the recursive-doubling (RD), SMP, and NAP methods for reducing a single value on each process for various process counts. 
Furthermore, Figure~\ref{figure:allreduce_scale_speedup} shows the associated speedups obtained with the NAP method.  
The NAP allreduce algorithm obtains notable speedups over the other methods, particularly at process counts that are a power of \texttt{ppn}.

\begin{figure}[ht!]
    \centering
    \includegraphics[width=\linewidth]{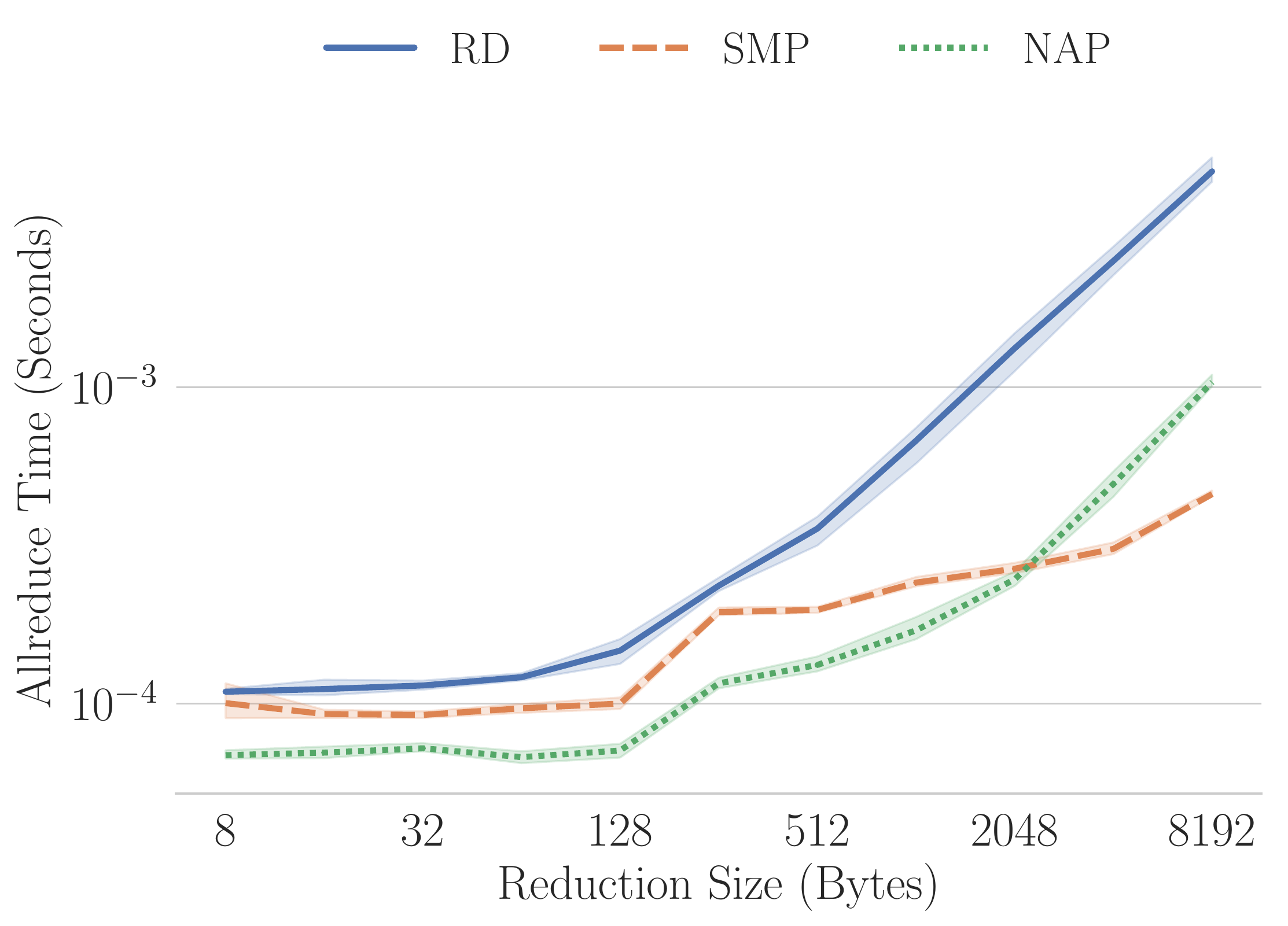}
    \caption{The cost of reducing various numbers of values over $32\,768$ processes with the recursive-doubling, SMP, and NAP allreduce methods.}\label{figure:allreduce_size_times}
\end{figure}

\begin{figure}[ht!]
    \includegraphics[width=\linewidth]{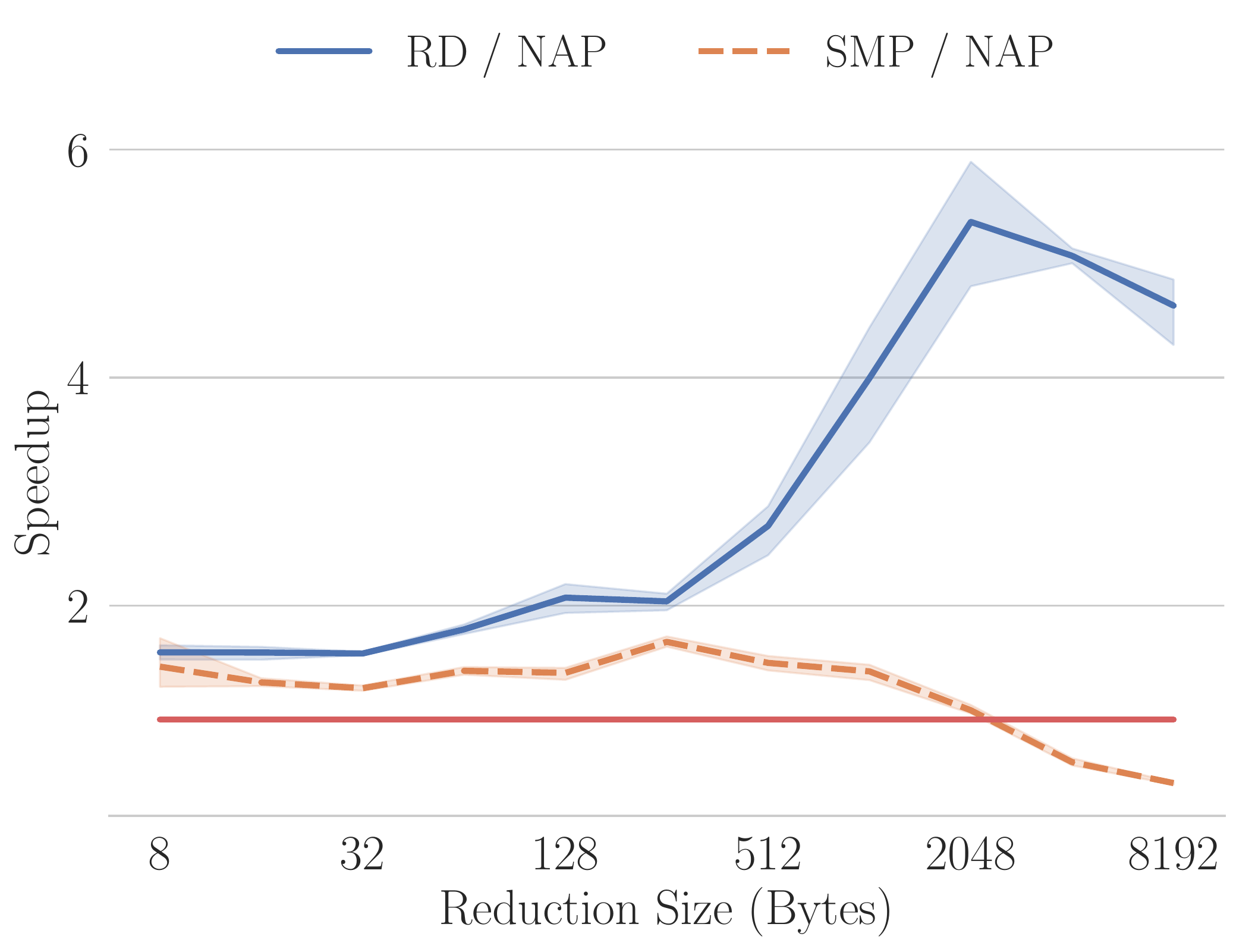}
    \caption{The speedup in the NAP allreduce algorithm over the recursive-doubling and SMP methods for reducing various numbers of values on $32\,768$ processes.  The NAP allreduce yields improved performance up to a reduction size of $2048$ bytes.}\label{figure:allreduce_size_speedup}
\end{figure}

Figures~\ref{figure:allreduce_size_times} and~\ref{figure:allreduce_size_speedup} show the costs and speedups, respectively, for performing the various allreduce methods on $32\,768$ processes for a variety of reduction sizes.  
The NAP method yields significant speedups over the recursive-doubling and SMP methods for smaller message sizes.  However, the SMP approach outperforms the NAP method for reduction sizes over $2048$ bytes, similar to expected performance based on the models in Figure~\ref{figure:perf_model_scale}.

\section{Conclusions and Future Work}\label{section:conclusion}
The NAP allreduce method yields notable improvements over standard recursive-doubling and existing node-aware SMP methods, in both performance models and measured costs for small message sizes, of up to $2048$ bytes.  The NAP algorithm relies on power of $\texttt{ppn}$ process counts, but natural extensions allow for all other process counts.  However, non power of $\texttt{ppn}$ process counts require the same number of inter-node communication steps as the succeeding power of $\texttt{ppn}$.  Therefore NAP allreduce speedups are most significant at power of $\texttt{ppn}$ process counts.
\begin{figure}[ht!]
    \centering
    \includegraphics[width=\linewidth]{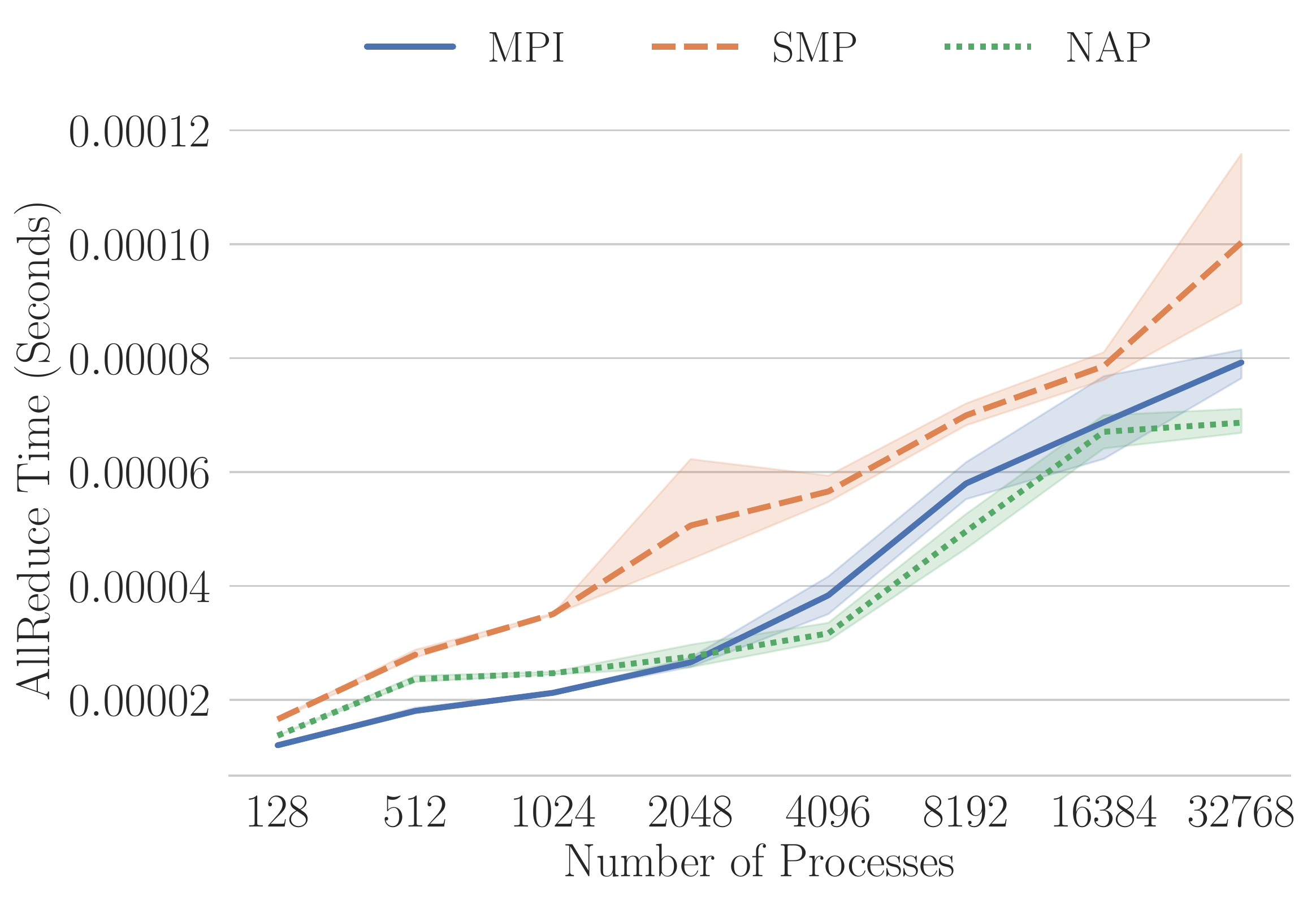}
    \caption{The cost of reducing various numbers of values over $32\,768$ processes with the SMP algorithm as implemented in MPICH (labeled MPI), SMP, and NAP allreduce methods.}\label{figure:compare_size_times}
\end{figure}
\begin{figure}[ht!]
    \includegraphics[width=\linewidth]{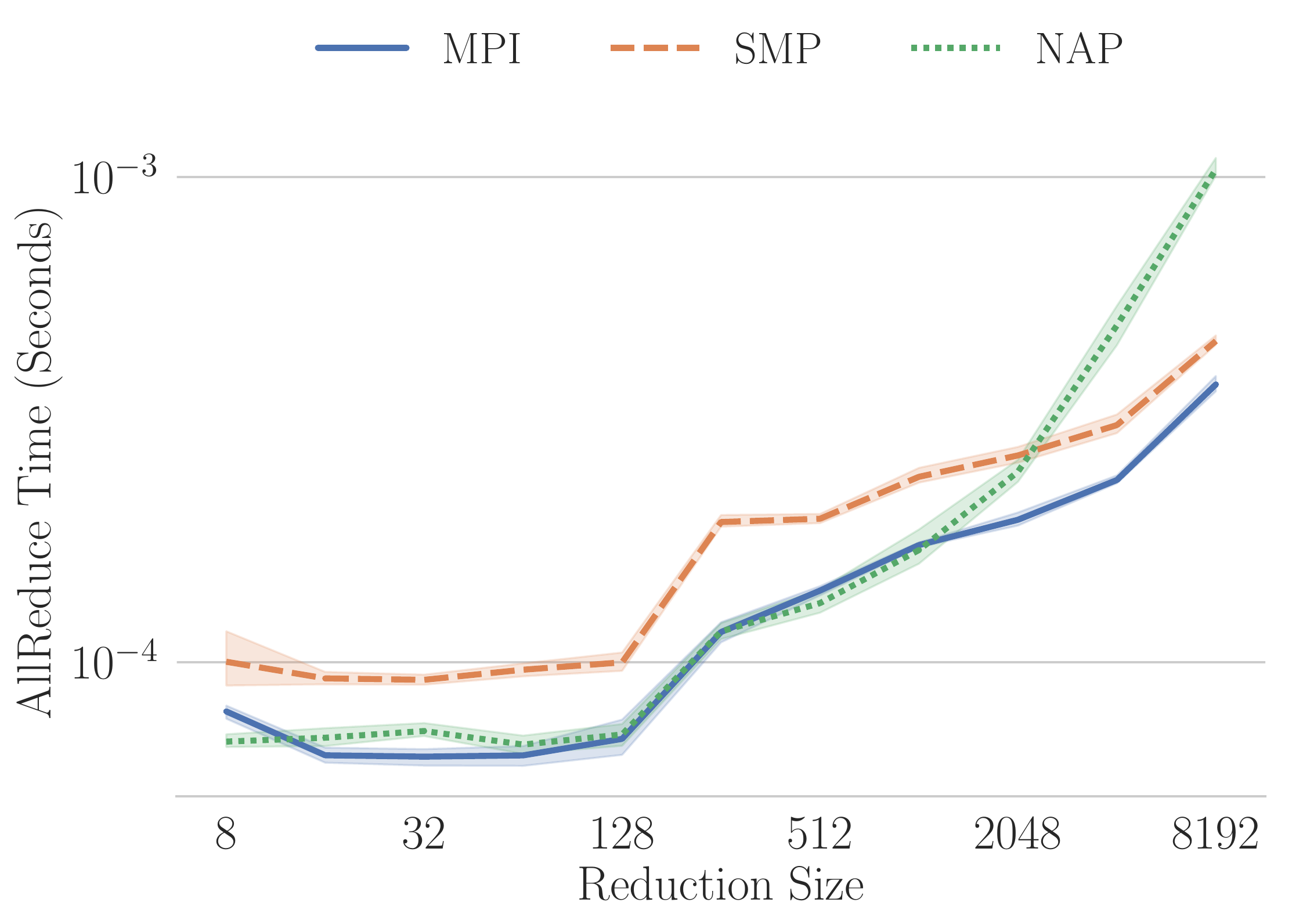}
    \caption{The cost of reducing various numbers of values over $32\,768$ processes with the SMP algorithm as implemented in MPICH (labeled MPI), SMP, and NAP allreduce methods.}\label{figure:compare_scale_times}
\end{figure}

This paper is focused on the cost of the recursive-doubling, SMP, and NAP allreduce methods when implemented on top of MPICH, calling \texttt{MPI\_Send} and \texttt{MPI\_Recv} for each step of communication.  However, there is significant overhead associated with these calls, in comparison to direct implementation of these methods in MPICH.  Figures~\ref{figure:compare_size_times} and~\ref{figure:compare_scale_times} display the cost of performing the SMP and NAP allreduce methods on top of MPI, compared to the SMP method as implemented in MPICH, measured by calling the \texttt{MPI\_Allreduce} routine.  The overhead associated with implementing on top of MPICH can be seen as the difference between the MPI and SMP costs.  While the node-aware allreduce yields slight improvements over MPICH's SMP approach, speedups are minimal due to the additional overhead.  Therefore, this method should be implemented as a part of MPICH to achieve optimal performance.

Similar node-aware approaches can be extended to other collective algorithms.  Natural extensions exist to the \texttt{MPI\_Allgather}, in which a similar recursive-doubling algorithm performs well for small gather sizes.  Furthermore, the node-agnostic ring algorithm again has multiple processes communicating duplicate data between nodes, which could be improved upon.  Using the max-rate model as a guide, node-aware extensions could be applied to larger \texttt{MPI\_Allreduce} methods, optimizing the reduce-scatter and allgather approach to avoid injection bandwidth limits while utilizing as many processes per node as possible.  

Finally, locality-aware collective algorithms can be extended to other parts of the architecture, such as reducing inter-socket communication in exchange for increased intra-socket message counts.  Similary, these algorithms can be optimized for heterogeneous architectures, in which many layers of memory and communication exist.

\bibliographystyle{IEEEtran}
\bibliography{IEEEabrv,refs}

\end{document}